\documentclass [twocolumn, prb, aps, showpacs] {revtex4}

\usepackage{graphicx}
\usepackage{amsmath}
\usepackage{epsfig}

\begin{document}

\title {Ferromagnetism in Carbon doped Zinc Oxide Systems}

\author{B. J. Nagare}
\email{bjnagare@physics.mu.ac.in}
\affiliation{
     Department of Physics,
     University of Mumbai, Santacruz (East),
     Mumbai-400 098,
     India.
}

\author {Sajeev Chacko}
\email{sajeev.chacko@gmail.com}

\author {D. G. Kanhere}
\email{kanhere@unipune.ernet.in}
\affiliation{
     Department of Physics and Center for Modeling and Simulation,
     University of Pune, Ganeshkhind,
     Pune - 411 007,
     India.
}
\begin{abstract}
We report spin polarized density functional calculations of ferromagnetic 
properties of a series of ZnO clusters and solid containing one or two 
substitutional carbon impurities. We analyze the eigen value spectra, 
spin densities and molecular orbitals, and induced magnetic moments for 
ZnC, Zn$_{2}$C, Zn$_{2}$OC, carbon substituted clusters Zn$_{n}$O$_{n}$ 
(n=3--10, 12) and ZnO solid. The results show that the doping induces magnetic 
moment of the $\sim$2~$\mu_{B}$ in all the cases. All the systems with 
two carbon impurities show ferromagnetic interaction except when the 
carbon atoms share the same Zn atom as the nearest neighbor. This ferromagnetic 
interaction is predominantly mediated via $\pi$ bonds in ring structures and 
through $\pi$ and $\sigma$ bonds in three dimensional structure. 
The calculations also show that the interaction is significantly enhanced 
in solid, bringing out the role of dimensionality of Zn-O network connecting 
two carbon atoms.

\end{abstract}


\maketitle

\section{Introduction}

Dilute magnetic semiconductors (DMS) are a focus of much attention due to 
their potential application as spintronics material.  These materials, apart 
from having a desirable band gap can exhibit room temperature ferromagnetism 
(RTF) upon doping by suitable impurities. Several oxides like TiO$_{2}$, 
SnO$_{2}$, In$_{2}$O$_{3}$, HfO$_{2}$ and more popular ZnO have been 
investigated and have been found to show ferromagnetism when doped with 
transition metal impurities.~\cite{Coey-natmater05, Hong-apl05, He-apl05, 
Hong-prb05, MVenkatesh-prl06}  Amongst these, ZnO is preferred because of 
many attractive features. It has a wide band gap of 3.37~eV, suitable for 
applications with short wavelength light, a large exciton binding energy, 
a low lasing threshold, and is transparent to visible light.

There are a number of theoretical and experimental studies  reported on ZnO 
yielding a variety of data on magnetic behavior. The data has raised many 
issues and at times, controversies. The density functional calculations 
show that doping by Cu produces half metallic ferromagnet~\cite{lin-prb06},
 the prediction has been verified experimentally.~\cite{buchholz-apl05}
Co doping has generated much controversy about the origin of magnetism. 
Cobalt doped ZnO is reported to have high Curie temperature (TC) of up to 700 K.  
A robust ferromagnetic state is predicted at the Oxygen surface even in the absence 
of magnetic atoms.~\cite{Sanch-prl08, Aron-prl08} If Mn or Co is doped 
in films of ZnO, ferromagnetic behavior has been observed in both insulating 
and metallic films, but not when the carrier density is intermediate.~\cite{behan-prl08}

Recently, non magnetic impurities have been shown to induce ferromagnetic 
behavior.  Vacancy-induced magnetism in ZnO thin films and nano wires 
have been reported on the basis of density functional investigations.~\cite{Wang-prb08} 
\begin{figure*}[h!t!p!]
\centering
\includegraphics[scale=0.34]{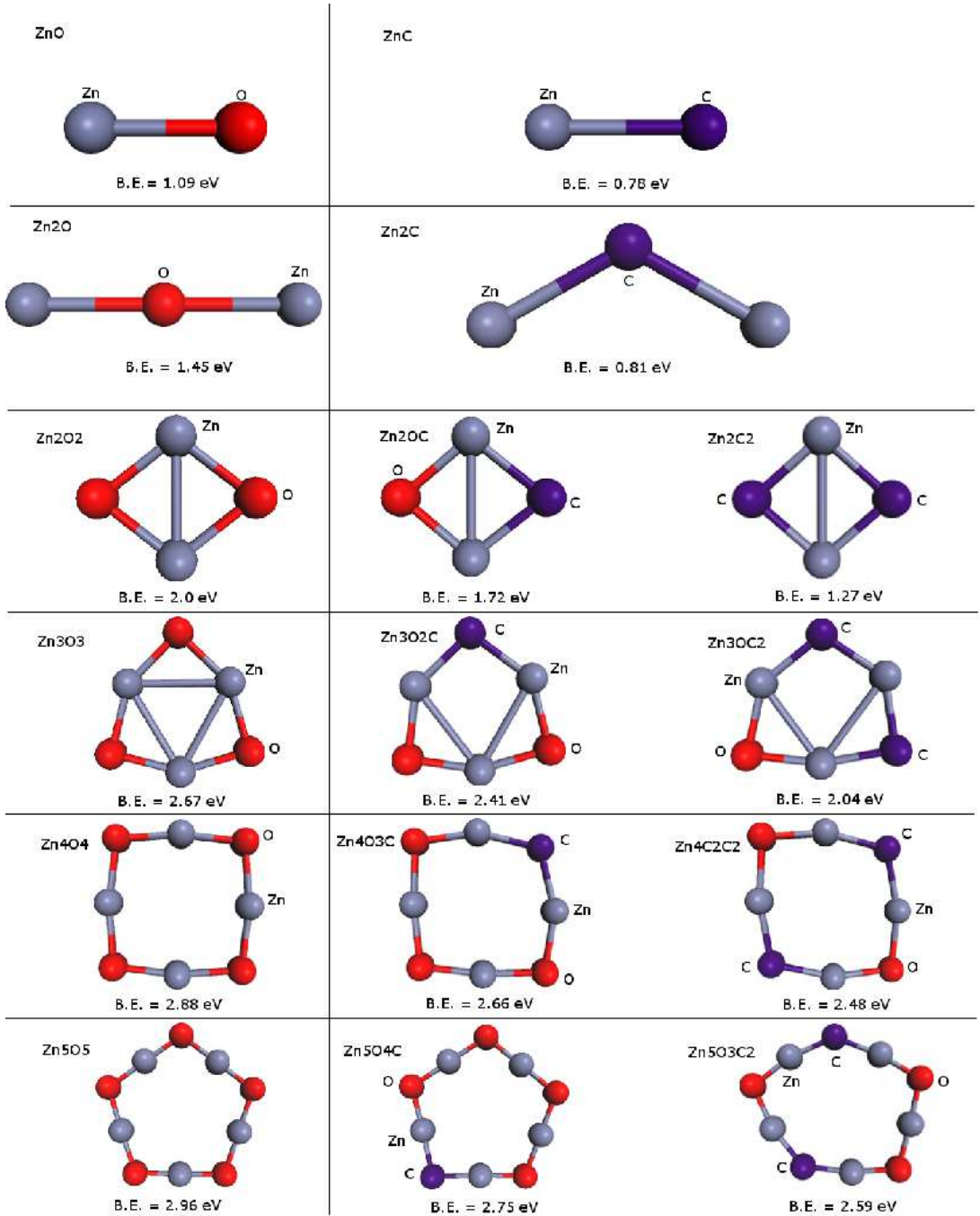}
\caption{Optimized ground state geometries of C-doped (ZnO)$_{n}$ (n=1-5)
clusters. The binding energy per atom (E$_{b}$) in eV is also shown.
The oxygen atoms are indicated by red color (online) whereas substituted
carbon is marked by C. The second and third column shows structures with
single and double carbon substitution respectively.}
\label{Figure 1}
\end{figure*}
In fact, defects of several types like vacancy at Zn site, oxygen surface, 
interstitial edges in nano ribbons~\cite{And-Nano08} etc. have been 
reported to be inducing magnetic interactions. Recently, carbon doped ZnO 
has been shown to exhibit ferromagnetic behavior with Curie temperature 
around room temperature.~\cite{Pan-prl07} The experiment has been performed 
with two levels of doping, 1~$\%$ and 2.5~$\%$. The spin density functional 
theory (SDFT) results show that the magnetism is due to ZnC system. 
The magnetic moment per carbon atom decreases slightly for the higher 
concentration from 2-3~$\mu_{B}$ to 1.5-2.5~$\mu_{B}$.
There are several unclear issues: the origin of magnetic 
moment on carbon site, the role of Zn and oxygen, the nature of the long 
range interaction between the two carbon sites, the role of oxygen--Zn network
 connecting the two sites. In order to gain some insight in to some of 
these issues, we have carried out a systematic density functional 
investigations on a series of systems, beginning with ZnC molecule. 
We have investigated the evolution of magnetic moments and magnetic 
interaction in ZnC, Zn$_{2}$C, Zn$_{2}$OC, Zn$_{2}$C$_{2}$ and then in a 
series of clusters Zn$_{n}$O$_{n}$ (n=3--12) with a single and double 
carbon substitution.
 Finally we examine ZnO solid with a large unit cell of 72 atoms. 
We bring out the role of exchange splitting in Carbon atom, hybridization of 
Carbon p orbitals with Zn, the role of oxygen--Zn network, especially the 
role of dimensionality in stabilizing magnetic interactions.   
\section{Computational Method}

The calculations have been performed using SDFT with ultra soft 
pseudopotential and plane wave basis. 
\cite{Van-prb90, Cla-zfk05} We have used two different codes namely 
CASTEP\cite{Cla-zfk05} and VASP.\cite{kresse-prb99} 
\begin{figure*}[h!t!p!]
\centering
\includegraphics[scale=0.34]{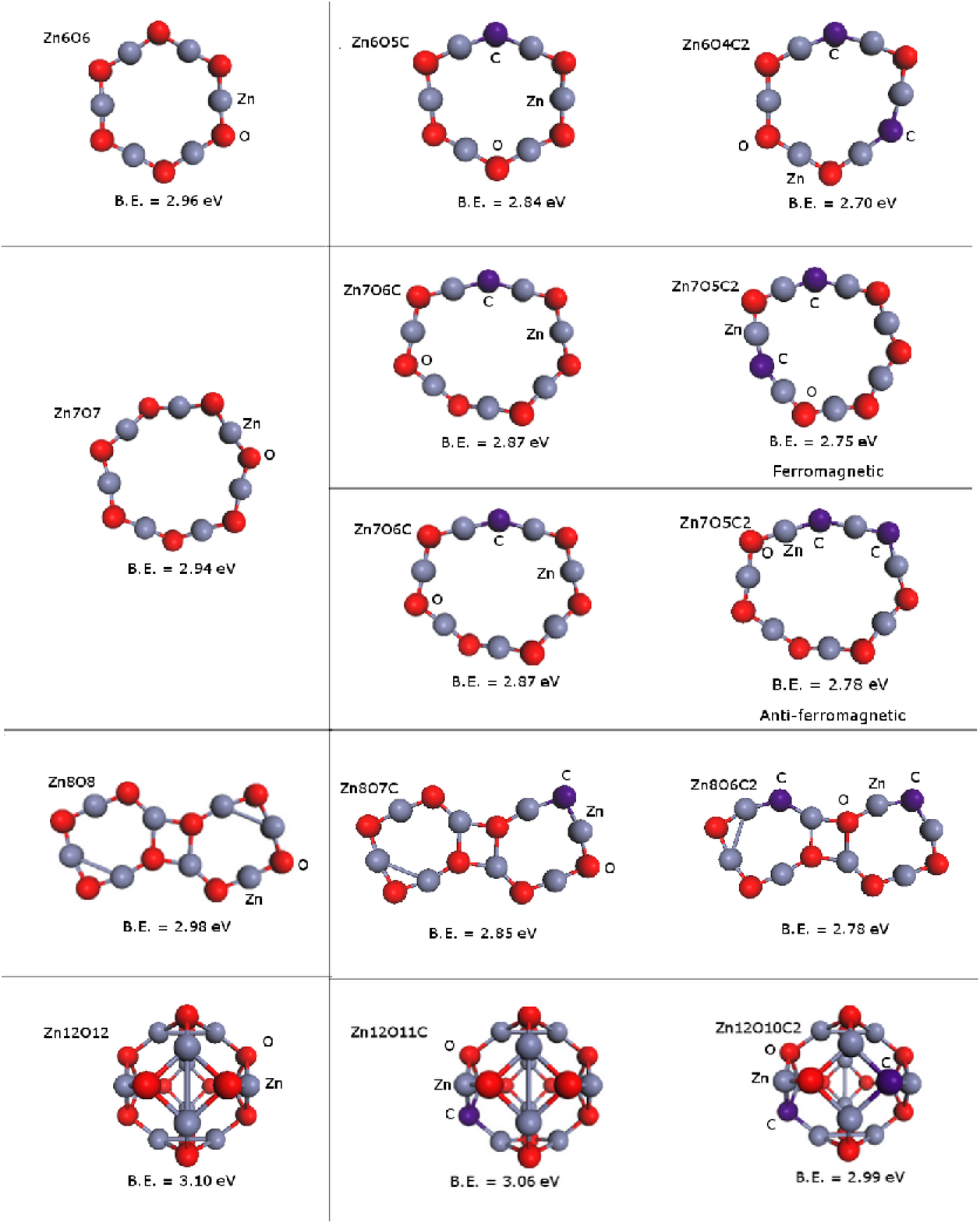}
\caption{Optimized ground state geometries of C-doped (ZnO)$_{n}$ (n=6-8, 12)
clusters. The binding energy per atom (E$_{b}$) in eV is also shown.
The oxygen atoms are indicated by red color (online) whereas substituted
carbon is marked by C. The second and third column shows structures with
single and double carbon substitution respectively.}
\label{Figure 2}
\end{figure*}
In all the cases, we have used generalized gradient approximation (GGA) 
given by Perdew-Burke-Ernzerhof (PBE).\cite{Perdew-prl96} 
For clusters we have used supercells having lengths $\sim$15--30~\AA. 
The energy cut-off for the plane wave basis was chosen to 300~eV. 
The structures were considered to be converged when the force on each ion 
was less than $\sim$0.05~eV/\AA~ with a maximum displacement of 
$\sim$0.002~\AA~ and a convergence in the total energy of about 
$2\times 10^{-6}$ eV/atom. 
In all the cases, the geometries of all the lowest energy structure are in 
good agreement with results of Matxain et. al.\cite{Mat-pra00} 
To study the magnetic properties of C-doped (ZnO)$_{n}$ clusters, carbon 
atoms were doped substitutionally at various O-sites. The C-doped clusters 
were also optimized with the same set of parameters as used for 
(ZnO)$_{n}$ clusters. 

We have verified the accuracy of our model on ZnO molecule and ZnO solid. 
We find a bond length of 1.736~\AA~ and a binding energy 1.79~eV for ZnO
molecule. The previous studies\cite{Gust-jpca00} also report a bond length 
between 1.71-1.75~\AA. We also calculate the binding energy of the ZnO 
solid in the wurtzite structure. The calculated binding energy of 7.23~eV 
per ZnO unit agrees well with the experimental binding energy of 7.52~eV 
per ZnO as reported by Jaffe et.al.\cite{Jaffe-prb00} 

In order to understand the basic mechanism of ferromagnetic origin and 
evolutionary trends, we have performed the calculations on series of 
clusters namely; ZnC, Zn$_{2}$C, (ZnO)$_{n}$ and the substitutionally 
doped Zn$_n$O$_{n-m}$C$_m$ clusters with $n=2-10,~12$ and $m=1-2$ as 
well as on ZnO solid. In each of the cases, we substitute either one or 
two carbon atoms at O--sites. 

In next section, we present and discuss our results.
\begin{figure}
\includegraphics[angle=270, scale=0.35]{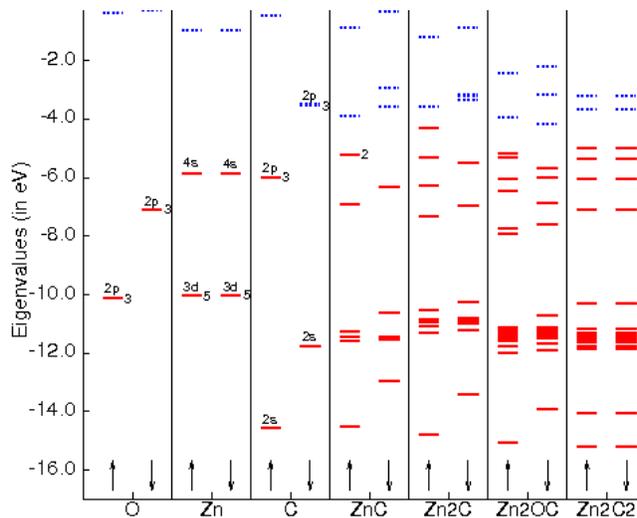}
\caption{The Eigenvalue spectra for some clusters. In each of the cases 
the left column is for up and right column is for down electrons. 
The unoccupied levels are shown by dotted lines. The numbers near the line 
shows degeneracy.}
\label{zno-fig-evs}
\end{figure}
\section{Results and Discussions}

It is most fruitful to examine the electronic structure and systematics of 
magnetic properties of some important motifs in zinc oxide solid.  
When the carbon atom is substitutionally doped at O-site in the bulk 
zinc oxide, it has zinc atom as the first nearest neighbor and oxygen 
atom as the second nearest neighbor. Therefore, we begin by examining 
the magnetism and the electronic structure of ZnC, Zn$_{2}$C, Zn$_{2}$OC, 
and Zn$_{2}$C$_{2}$ clusters. We have also calculated the electronic 
structure and the magnetic properties of ring structures of Zn$_{n}$O$_{n}$ 
cluster (n=3--8) and Zn$_{9}$O$_{9}$--Zn$_{12}$O$_{12}$ which form three 
dimensional structures. The results are also presented for bulk zinc 
oxide by using large unit cell with 72 atoms. The optimized ground state 
geometries of these ZnO systems are shown in figure 1 and 2. It can be seen 
that the substitutional doping by carbon atoms has a small effect on the 
geometries of the hosts with exception of Zn$_{2}$O.
\begin{table*}
\caption {The magnetic moment $\left(\mu_{M}\right)$ in $\mu_{B}$, 
Exchange energy E$_{J}$ (Energy difference between ferromagnetic and 
antiferromagnetic state) in meV, the nature of magnetic ground state for 
all the systems investigated.}
\begin{tabular}{|c|c|c|c|c|c|}
\hline\hline
Host Systems & \multicolumn{2}{c|} {Magnetic Moment $\mu_M$} & E$_J$ (meV) & Nature of & Geometry \\
\cline{2-3}
& Singly & Doubly & & ground state & \\
& doped  & doped  & &              & \\
\hline\hline

C                  & 2 & - & -     & -     & -                    \\
ZnO                & 2 & - & -     & -     & Linear               \\
Zn$_{2}$O$_{2}$    & 2 & 0 & -     & AFM   & Rhombous             \\
Zn$_{3}$O$_{3}$    & 2 & 0 & -     & AFM   & Cap-triangle         \\
Zn$_{4}$O$_{4}$    & 2 & 4 & -26.4 & FM    & Square               \\
Zn$_{5}$O$_{5}$    & 2 & 4 & -3.3  & FM    & Ring                 \\
Zn$_{6}$O$_{6}$    & 2 & 4 & -2.5  & FM    & Ring                 \\
Zn$_{7}$O$_{7}$    & 2 & 4 & -1.1  & FM    & Ring                 \\
Zn$_{7}$O$_{7}$~\footnote{Zinc is nearest neighbour to both carbon atoms.}
& 2 & - & -     & AFM   & Ring                 \\
Zn$_{8}$O$_{8}$    & 2 & 4 & -0.9  & FM    & Two connected rings  \\
Zn$_{9}$O$_{9}$    & 2 & 4 & -46.3 & FM    & 3D                   \\
Zn$_{10}$O$_{10}$  & 2 & 4 & -26.3 & FM    & 3D                   \\
Zn$_{12}$O$_{12}$  & 2 & 4 & -85.3 & FM    & 3D                   \\
ZnO soild~\footnote{Calculation was performed on ZnO solid within large unit
cell containing 72 atoms.}
& 2 & 4 & -89.3 & FM & Wurtzite \\ [6pt]
\hline\hline
\end{tabular}
\label{zno-table-frag}
\end{table*}
\begin{figure*}
\includegraphics[scale=0.25]{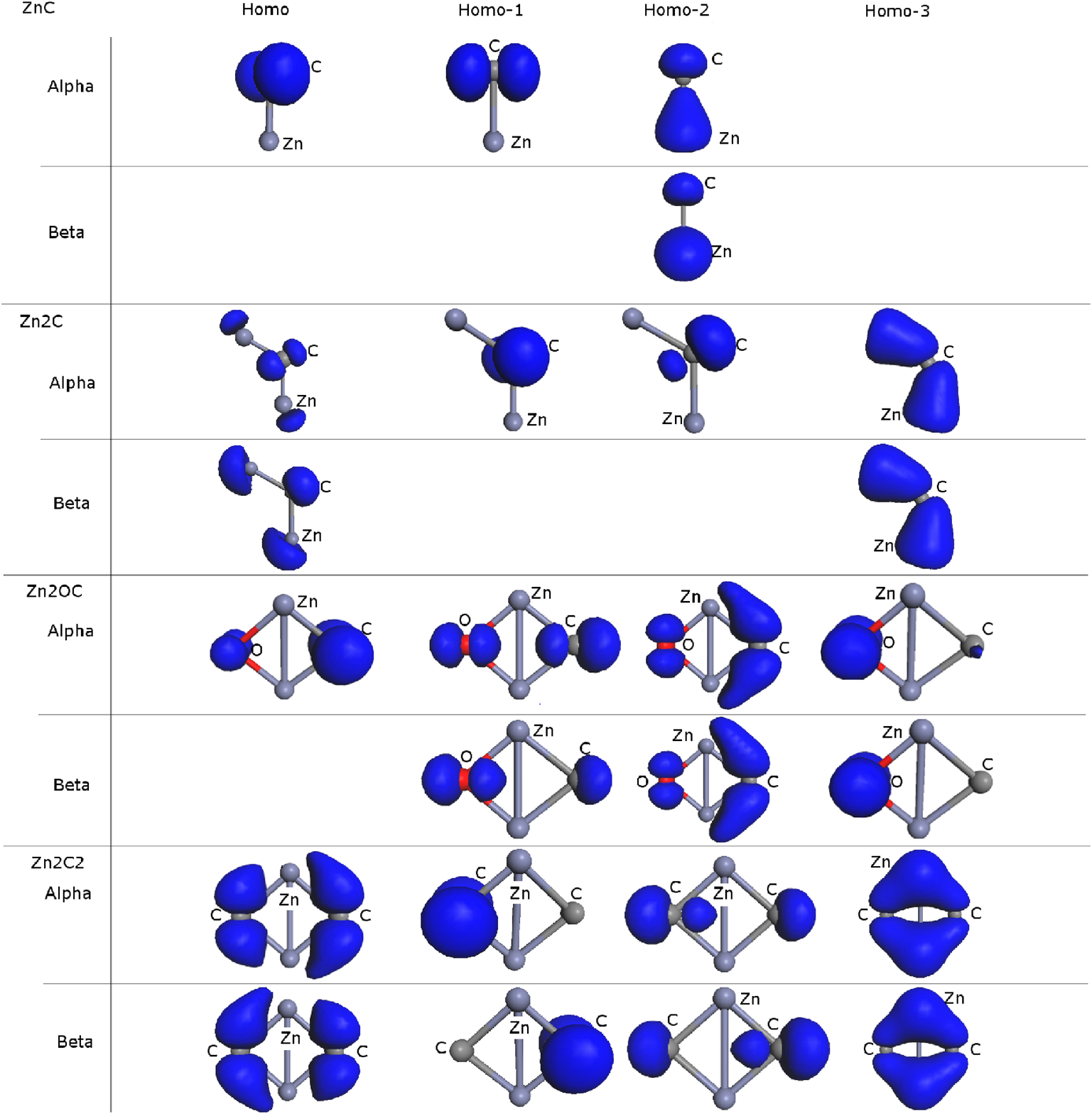}
\caption{Isovalued surfaces of molecular orbitals for up(alpha) and down(beta)
 electrons. The position of substituted carbon is marked by C.}
\label{zno-fig-MO}
\end{figure*}

In table 1, we summerize the results of all C-doped clusters. The table 
shows the magnetic moments per site, the exchange energy E$_{J}$ calculated 
as E$_{FM}$-E$_{AFM}$ (wherever appropriate) where, E$_{FM}$ and E$_{AFM}$ 
are the total energies of ferromagnetic and antiferromagnetic state 
respectively, the nature of magnetic ground state and nature of ground 
state geometries of C-doped clusters. A number of features can be discerned 
from table 1. 1. A majority of doubly doped (ZnO)$_{n}$ clusters turn out to 
be ferromagnetic with magnetic moment of $\sim$2~$\mu_{B}$ per carbon atom.
2. Whenever the same zinc atom is the nearest neigbhor to both the carbon 
atoms, the ground state is antiferromagnetic. This is the case for 
Zn$_{2}$C$_{2}$, Zn$_{3}$OC$_{2}$ and in other clusters when the carbon 
atoms are substituted on both sides of zinc atom. 
3. It turns out that $\sim$80-90~\% magnetic moment is localized on carbon sites
 with small but significant induced magnetic moment on oxygen and zinc sites 
(see figure 5 and the discussion). This is true for all singly and doubly 
substituted clusters irrespective of the placement of carbon atoms. 
4. A remarkable observation is that the exchange energy (E$_{J}$) is sensitive 
to the dimensionality of clusters and is highest for three dimensional 
clusters. Thus, the number of paths connecting two carbon atoms via 
zinc-oxygen network is a significant factor favouring the ferromagnetic 
coupling. In a sense this also brings out the role of oxygen in establishing 
the ferromagnetic interaction. We elaborate on this point in the discussion 
below.  

In order to elucidate the origin of magnetism, we examine the eigenvalue 
spectra (figure 3) and the nature of molecular orbitals for ZnC, Zn$_{2}$C, 
Zn$_{2}$OC and Zn$_{2}$C$_{2}$ along with atomic Zn, O, and C which are shown 
in figure 4. We use the nomenclature alpha homo and beta homo to denote the 
highest occupied molecular orbitals for up spin and down spin respectively. 
In figure 3, we show the eigenvalue spectra for up spin (left part) and down 
spin (right part) electrons. The relevant molecular orbitals are shown in 
figure 4 as isovalued surfaces. We also show the induced magnetic moments 
on all sites in figure 5 for some of the relevant systems. We will discuss 
the spectra, molecular orbitals and site distribution of the magnetic moments 
together. Let us note that the exchange spliting for carbon atom between 
spin up and spin down electron is $\sim$2.49~eV. The magnetic moment is due 
to two up spin occupied orbitals, consistent with the Hund's rule. This brings 
the eigen values of atomic carbon p and zinc 4s quite close to each other for 
up states, while the down spin eigen value of carbon p is placed significantly
 higher. The formation of the spectra across the series reveals that the 
action is in the s-p complex formed by carbon p and zinc 4s orbitals, and the 
spliting of the up and down "bands" persists for all the systems, leading 
to the formation of the magnetic moment around carbon. 
\begin{figure*}
\includegraphics[scale=0.36]{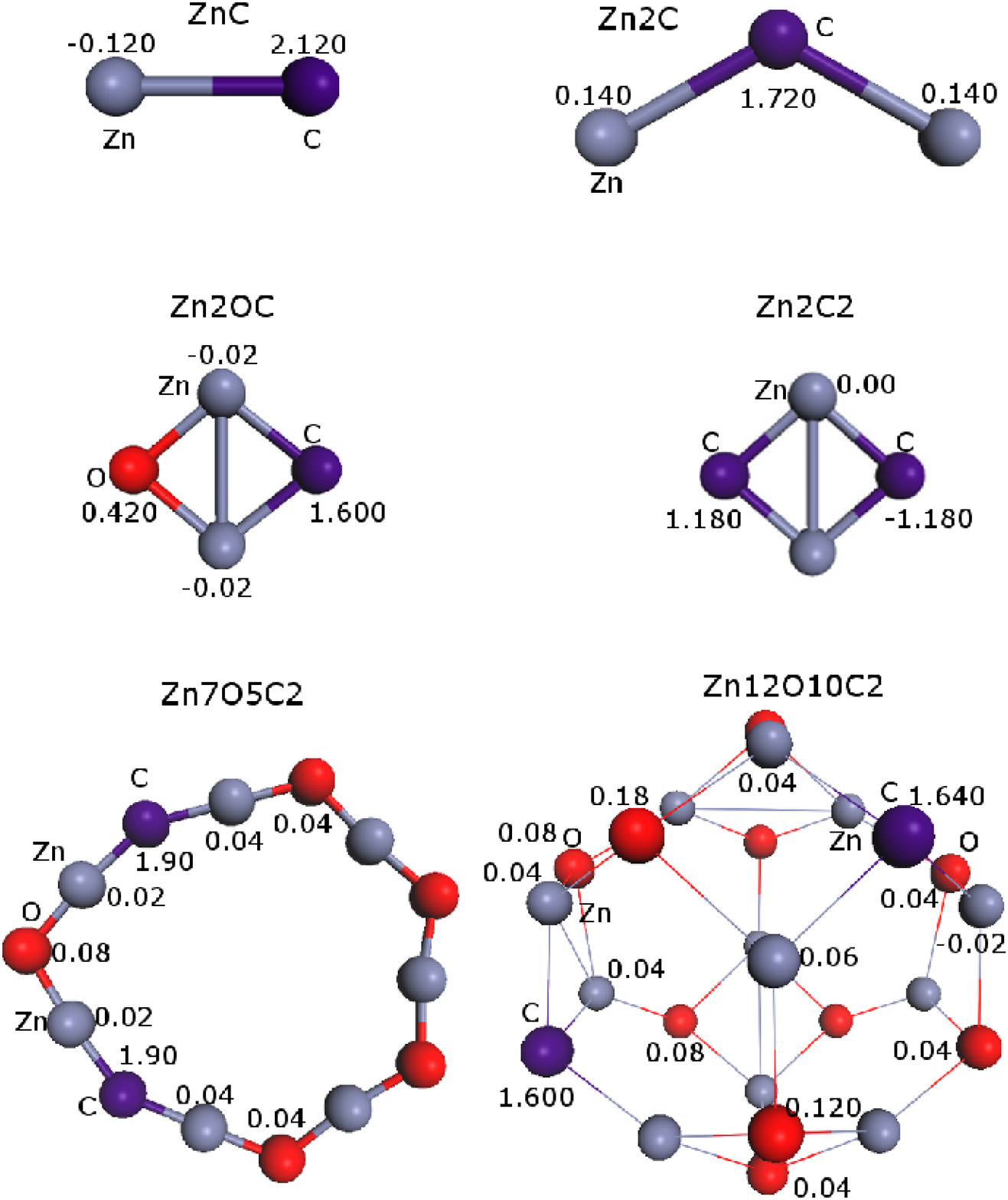}
\caption{Spin moment at site is depicted in C-doped clusters. The position of
substituted carbon is marked as C.}
\end{figure*}

ZnC carries a magnetic moment of $\sim$2~$\mu_{B}$ on carbon site. Because 
of the linear nature of ZnC, only one of the p orbitals is capable of 
hybridizing with Zn s. As a consequence triply degenerate p states (for the 
both the spins) split into two states as a singly degenerate and doubly 
degenerate once, retaining the exchange splitting between up and down. 
The corresponding molecular orbitals are shown in figure 4. A lowest state in 
the complex for both the spins are of similiar nature and show significant 
hybridization and delocalization. The magentic moment is due to the doubly 
degenerate pure p states localized on the carbon atom. The Mulliken population 
analysis shows effective charge transfer from Zn to C of order of $\sim$1.7 
electrons. Thus ignoring the splitting between up-down eigenvalues, a simple 
picture emerges. The charge transfer from zinc makes one of p orbitals on 
carbon doubly occupied. The delocalization caused by hydridization with 4s 
reduces Coulomb repulsion. The magentic moment is due to the doubly degenerate 
pure p states localized on the carbon atom. 

\begin{figure*}
\includegraphics[scale=0.36]{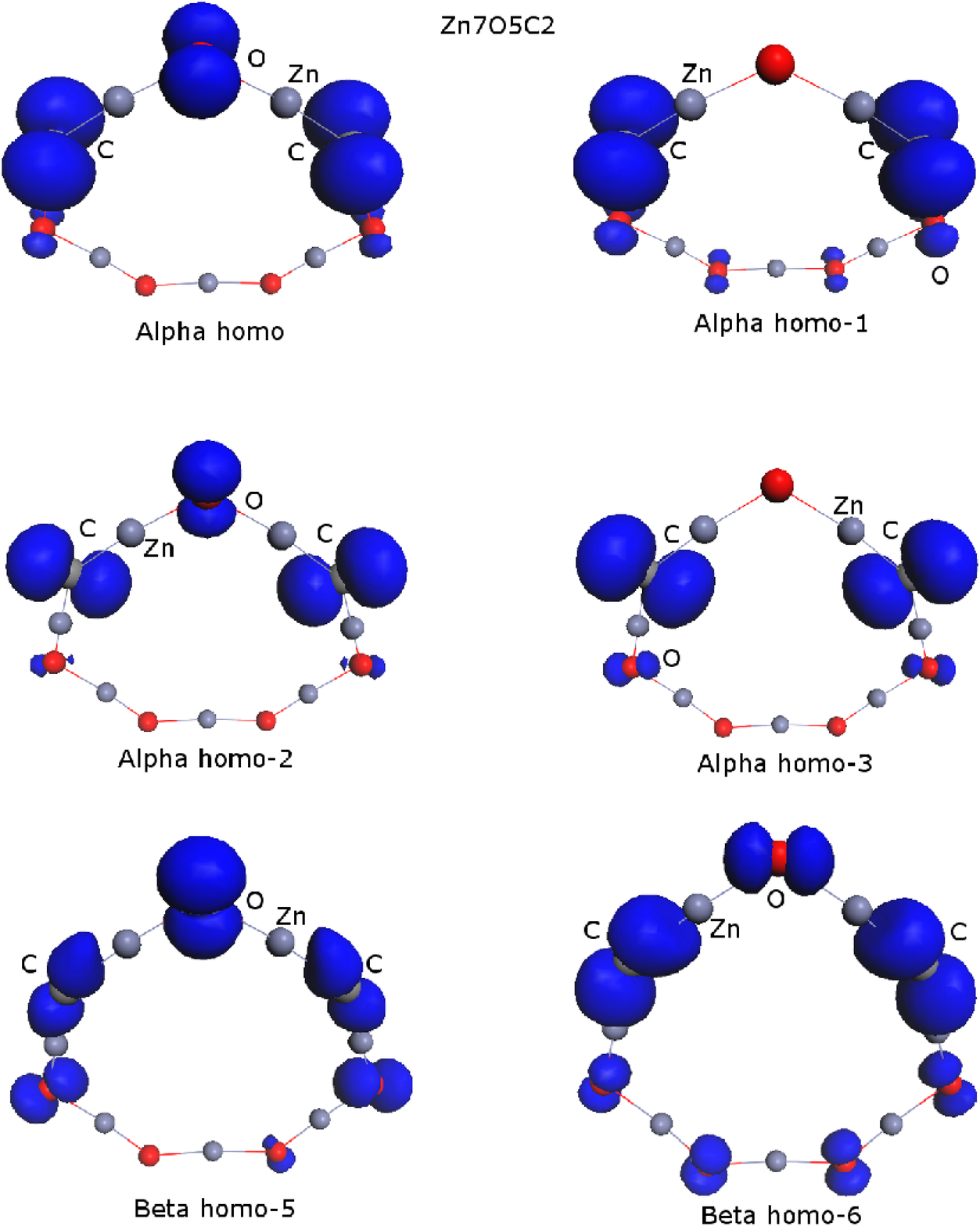}
\caption{Isovalued surfaces of molecular orbitals for up and down elecrons
in C-doped Zn$_{7}$O$_{7}$ clusters. The oxygen atoms are indicated by
red color (online) whereas the position of substituted carbon is
marked by C.}
\end{figure*}

Next we turn to Zn$_{2}$C. Since in ZnC, two spin down orbitals on carbon 
are still unoccupied, one would naively expect Zn$_{2}$C to be nonmagnetic. 
Interestingly, Zn$_{2}$C also shows a magnetic moment of $\sim$2$\mu_{B}$ 
mainly localized around the carbon atom. The hybridization of one of the p 
orbitals with two zinc 4s splits the spectrum of up and down electrons 
forming bonding and antibonding orbitals, the splitting being much less 
in the case of down electrons. As a consequence four up and two down orbitals 
are occupied. Once again by examining the molecular orbitals (figure 4), we 
arrived at a simple description namely; there are two doubly occupied 
orbitals which are delocalized (2p-4s hybridized). The magnetic moment arises 
out of two carbon based p's orbitals (homo-1 and homo-2 for up electrons). 
As a consequence of the hybridization a small but significant magnetic moment 
is induced on both zinc atoms. 

\begin{figure*}
\includegraphics[scale=0.41]{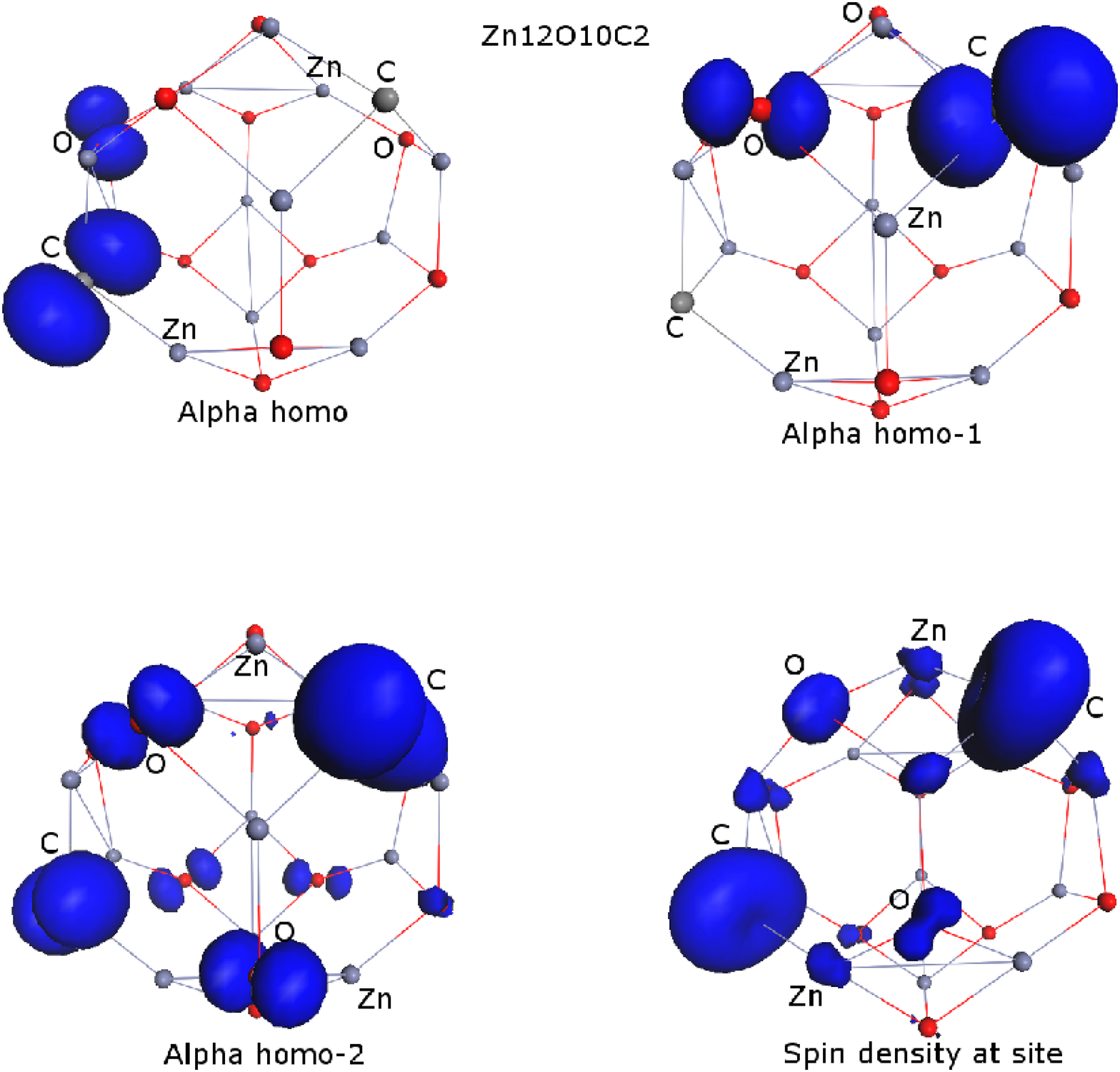}
\caption{Isovalued surfaces of molecular orbitals for up and down electrons
in C-doped Zn$_{12}$O$_{12}$ clusters. The isosurface spin density at site
is also shown. The position of substituted carbon is marked by C.}
\end{figure*}

Next we examine a typical motif Zn$_{2}$OC which brings in the role of oxygen. 
With introduction of oxygen, there is a competition between oxygen and carbon 
for charge transfer from zinc. The occupied molecular orbitals show 
interesting pattern. Evidently, relevant orbitals leading to magnetic 
interactions are homo and homo-1. The homo shows $\pi$ bond formation between
 carbon and oxygen (homo perpendicular to plane) while homo-1 shows 
the $\sigma$ bond. There are three down orbitals nearly compensating the 
charge in the three up orbitals. It can be seen that a significant magnetic 
moment ($\sim$0.42~$\mu_{B}$) is induced on the oxygen sites, clearly 
because of homo and homo-1. The existence of $\pi$ and $\sigma$ bonds 
establishes the coupling between carbon and oxygen and is a crucial 
element in the formation of long range magnetic interactions even in the 
extended systems. 

The ground state of symmetric cluster Zn$_{2}$C$_{2}$ is antiferromagnetic. 
There is a charge transfer of one electron to carbon atoms 
from each of the zinc atoms. As a consequence the zinc-zinc distance 
increases by $\sim$0.2~\AA. As can be seen from figure 4, it is homo-2 which 
is responsible for antiferromagnetic coupling. This is analog of super-exchange
 mechanism. It may be noted that a similiar observation can be made for 
Zn$_{3}$OC$_{2}$ as well as for all the ring clusters when carbon atoms 
are placed on both sides of zinc atom. We recall that the experiment indicate
a saturation of magnetic moment per carbon atom around 5~\% substitution.

\begin{figure*}
\includegraphics[scale=0.33]{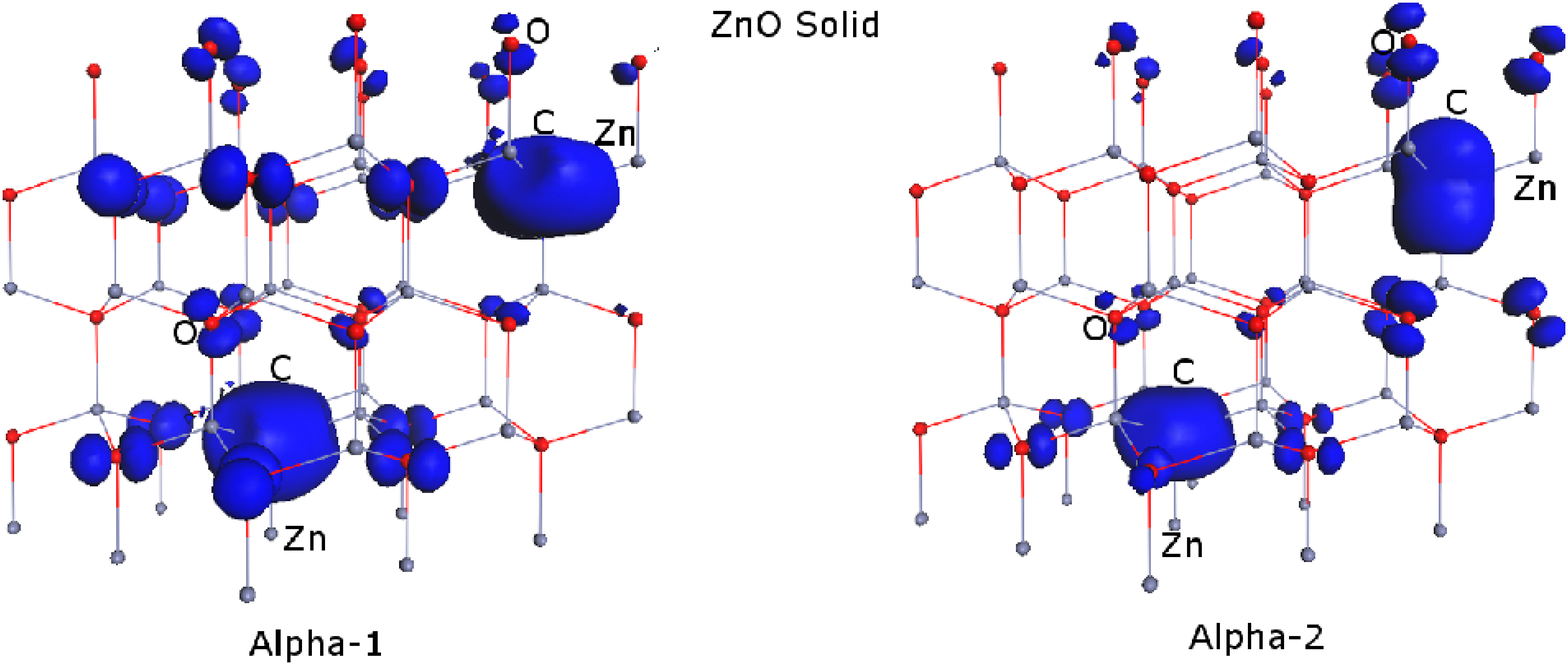}
\caption{Isovalued surfaces of molecular orbitals in C-doped zinc oxide solid 
for up electrons at the top of the valence band. The position of substituted 
carbon is marked as C.}
\end{figure*}

Now, we turn our attentation to ring structures Zn$_{5}$O$_{5}$ --
Zn$_{7}$O$_{7}$. We discuss the typical case of Zn$_{7}$O$_{5}$C$_{2}$ 
showing the ferromagnetic behaviour. The relevant molecular orbitals are 
shown in figure 6. In this case two carbon atoms are separated by chain 
of zinc-oxygen-zinc (see figure 5). It can be seen from alpha homo and 
alpha homo-1, that there is a $\pi$ bond formation between three p orbitals 
centered on two carbon atoms and one oxygen atom. A similiar bonding is 
also seen in alpha homo-2 and alpha homo-3 as well as in beta orbitals which
are shown in figure 6. These are the orbitals which are responsible for the 
ferromagnetic interaction between the carbon atoms mediated via oxygen.
\begin{figure*}
\includegraphics[scale=0.39]{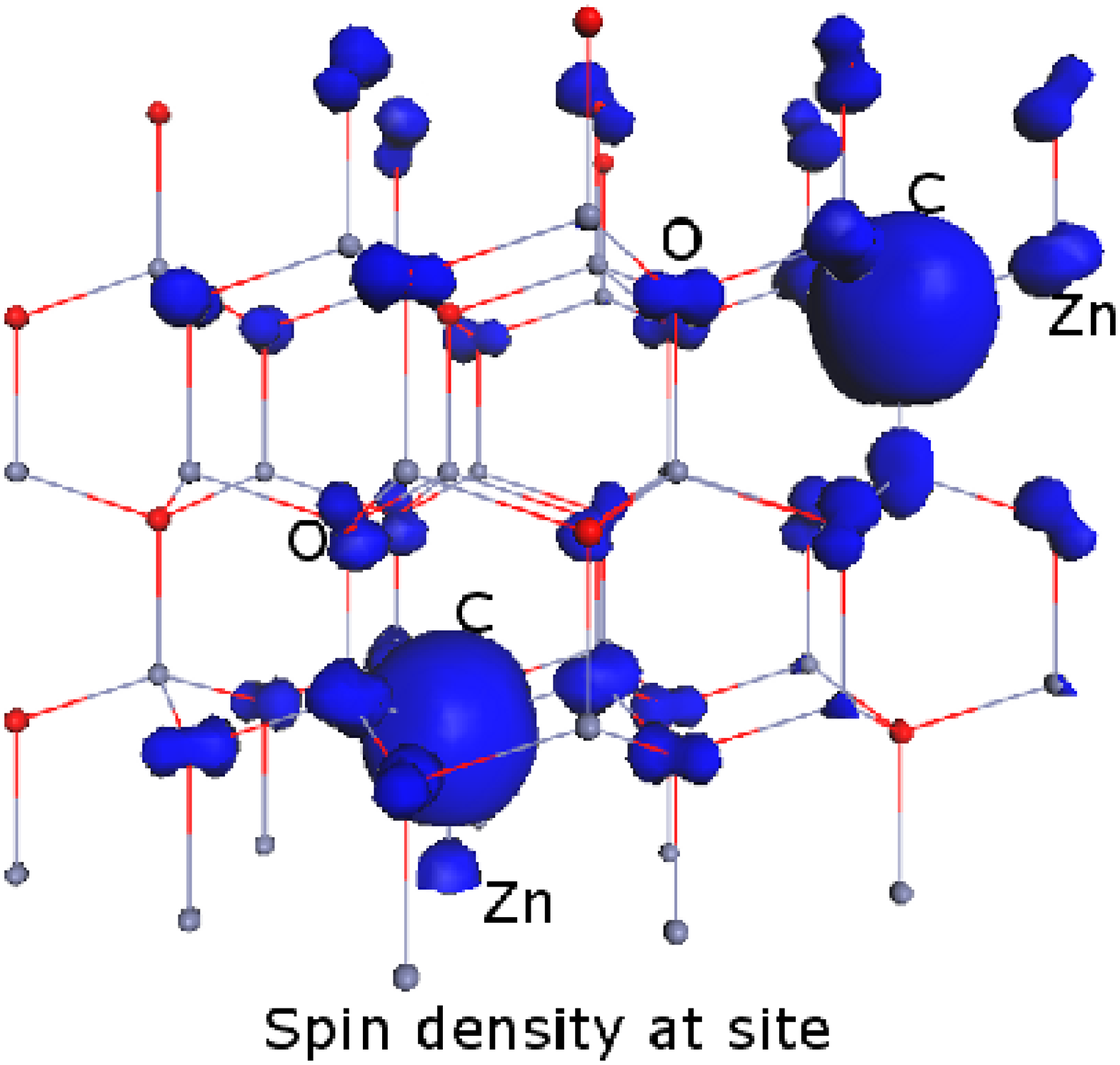}
\caption{The isosurface spin density at site is shown for the case of
C-doped zinc oxide solid. The position of substituted carbon is marked as C.}
\end{figure*}

Zn$_{12}$O$_{10}$C$_{2}$ is a three dimensional structure. As pointed out 
earlier when the structure changes to three dimensional one, there is a 
significant change in the connectivity of two carbon atoms via zinc and 
oxygen. The ground state structure for this system is shown in figure 2. 
The induced magnetic moments are shown in figure 5. It can be seen that the 
induced magnetic moments on two nearest neighbor oxygen sites are substancial 
$\sim$0.180~$\mu_{B}$. Other oxygen sites connecting them also have notable 
magnetic moments. The relavant molecular orbitals (alpha homo - alpha homo-2 ) 
and the spin density are shown in figure 7. Clearly the picture described in 
the earlier cases continues in the present case. 
The induced spin density shows the magnetic moment on carbon and oxygen sites. 
These homos show interaction between two carbon atoms mediated via oxygen. 
Thus, in these case once again the magnetic interaction between the carbon 
atoms is mediated via p orbitals ($\pi$-bond). The most significant point is 
that the interaction is strenghtened by availability of multiple sites 
connecting two carbon atoms. In this case, the C--C distance is 
$\sim$~5.410~\AA~. 

We expect a similiar picture to be valid in ZnO solid. We have considered a 
large unit cell with 72 atoms containing two carbon atoms separated by 7.6~\AA. 
Our results for the density of states (DOS) and magnetic moments are 
consistent with the report of of Pan et.al.\cite{Pan-prl07} We find that the 
spin-up bands are fully occupied while the spin-down bands are partially 
filled resulting in the magnetic moment of $\sim$2$\mu_{B}$ per carbon atom. 
We show the relavant molecular orbitals near the top of the valence 
band for up electrons. It can be noted from figure 8 that the two carbon 
atoms are separated by a zinc oxide plane. We also show the spin density in 
figure 9. The molecular orbitals depict two different ways in which the 
interaction is mediated. In the case of alpha-1, the carbon (in the top plane) 
induces magnetic moments on oxygen sites within the plane which then 
couples to the oxygen in the bottom plane as seen in figure 8. In the case 
of orbitals marked as alpha-2, the carbon induces the magnetic moments in the 
plane, below and above which couples the carbon site in the bottom plane. 
These figure clearly bring out the role of three dimensional oxygen network 
in providing the multiple paths between two carbon atoms. The resulting spin 
density is shown in figure 9. As expected the spin density is positive and is
 mainly around carbon sites. The induced magnetic moments on oxygen sites 
are clearly seen.   
\section{Summary and Concluding Remarks}

We have carried out spin polarized electronic structure calculations on 
series of (ZnO)$_{n}$ and the substitutionally doped Zn$_n$O$_{n-m}$C$_m$ 
clusters with $n=1-10,~12$ and $m=1-2$ as well as for zinc oxide solid with 
a view to elucidate the origin of magnetism in carbon doped zinc oxide. 
Our calculations reveal the crucial role played by oxygen in mediating 
ferromagnetic interaction. A detailed study of molecular orbitals show 
that the interaction is mediated via $\pi$ as well $\sigma$ bonds
 (especially in three dimensional solid) formed out of p electrons 
of carbon and oxygen. Our calculations also 
bring out the role played by hybridization of zinc 4s with carbon 2p 
and oxygen 2p. We expect the antiferromagnetic coupling if carbon 
atoms get substituted on sites adjacent to zinc leading to saturation
of magnetic moment with increasing carbon content. The results clearly 
show a significant enhancement in the ferromagnetic 
interaction between the two carbon atoms due to higher dimensionality, 
bringing out the role of multiple connectivity (between two carbon atoms) 
of the Zn-O network.

\begin{acknowledgments}
We are greatful to Accelrys Inc. USA for providing us the demonstration 
version of CASTEP. We also thanks to CDAC, India for providing us the high 
performance computing facility.  B. J. Nagare thanks to Balchandra Pujari 
and Vaibhav Kaware for their help during the project. 
\end{acknowledgments}

\bibliography{bibliography}
\end{document}